\documentclass[twocol]{ametsocV5}
\bibpunct{(}{)}{;}{a}{}{,}

\usepackage{hyperref}
\hypersetup{colorlinks=true, urlcolor=blue, linkcolor=blue, citecolor=blue}

\usepackage{amsmath,amsfonts,amssymb,bm}
\usepackage{mathptmx}
\usepackage{newtxtext}
\usepackage{newtxmath}

\title{Assimilation of the SCATSAR-SWI with SURFEX: Impact of local observation errors in Austria}

\authors{Jasmin Vural}
\affiliation{Numerical Weather Prediction, Zentralanstalt f\"ur Meteorologie und Geodynamik, Vienna, Austria}

\extraaffil{Numerical Weather Prediction, Zentralanstalt f\"ur Meteorologie und Geodynamik, Vienna, Austria}
\extraauthor{Stefan Schneider
}
\extraaffil{Numerical Weather Prediction, Zentralanstalt f\"ur Meteorologie und Geodynamik, Vienna, Austria}
\extraauthor{Bernhard Bauer-Marschallinger}
\extraaffil{Department of Geodesy and Geoinformation, TU Wien, Vienna, Austria}
\extraauthor{Klaus Haslinger}
\extraaffil{Climate Research, Zentralanstalt f\"ur Meteorologie und Geodynamik, Vienna, Austria}
\abstract{
The proper determination of soil moisture on different scales is  important for applications in a variety of fields. 
We aim to develop a high-level soil moisture product with high temporal and spatial resolution by assimilating the multilayer soil moisture product SCATSAR-SWI (Scatterometer Synthetic Aperture Radar Soil Water Index) into the surface model SURFEX. In addition, we probe the impact of the findings on the Numerical Weather Prediction (NWP) in Austria.
The data assimilation system consists of the NWP model AROME and the SURFEX Offline Data Assimilation, which provide atmospheric forcing and soil moisture fields as mutual input.
To address the known sensitivity of the employed simplified Extended Kalman Filter to the specification of errors, we compute the observation error variances of the SCATSAR-SWI locally using Triple Collocation Analysis and implement them into the assimilation system. 
The verification of the forecasted 2~m temperature and relative humidity against measurements of Austrian weather stations shows that the actual impact of the local error approach on the atmospheric forecast is slightly positive to neutral compared to the standard error approach, depending on the time of the year.
The direct verification of the soil moisture analysis against a gridded water balance product reveals a degradation of the unbiased root mean square error for small observation errors.
}

\begin{document}
\maketitle

\section{Introduction}
Soil moisture is an essential part of the energy-water cycle. Therefore, the availability of soil moisture products is important for many applications in, e.g., meteorology and hydrology. Especially when the incoming radiation is high, soil moisture can have a large impact on the lower atmosphere due to evapotranspiration and the involved partitioning into latent and sensible heat. 
As a consequence, forecast models used in NWP clearly profit from the initialisation using soil moisture fields \citep{2001hes,2002seugro,2007druvit,2014schwan,2016dirhal,2019drarei}.

In particular, high-resolution remotely sensed data provide new possibilities to assimilate soil moisture consistently on local to global scales \citep{2007dru,2008scidru,2010mah}. 
The potential of assimilating remotely sensed surface and multi-layer soil moisture was  successfully demonstrated by \citet{2014parmah} and \citet{2017albmun}.

The assimilation of near-surface variables has been shown to work well with a (simplified) extended Kalman Filter ([s]EKF; \citealt{2001hes,2009mahber,2011dramah,2013dedru}), which is also used in our study. 
Paramount to the successful application of a Kalman filter is the proper specification of observation and background errors for the Kalman gain. The Kalman gain is essential for the data assimilation as it determines the effect of the observation on the analysis with respect to the background. In the sEKF, observation and background errors are often set to the same value for simplicity, taking observations and background into account by an equal amount in the increment.

To address the known sensitivity of the Kalman filter to the error specification, constant (global) observation errors are sometimes scaled on the whole domain or grid-pointwise (locally) to find the optimum value \citep{2007pauver,2011barcal,2012drarei,2016derei}. For example, \citet{2019drarei} estimated the observation error by scaling the standard deviation of the soil-moisture time series for each grid-point to match previously found error distributions. In addition, observation errors can be assessed locally with Triple Collocation Analysis (TCA; \citealt{1993ros,1998sto}). Error estimates from TCA have already been utilized in (ensemble Kalman filter-based) data assimilation systems \citep{2010crovan,2014croyil,2015grucro}. 

In this study, we want to investigate the benefits of (1) assimilating the daily multi-layer 1~km soil moisture product SCATSAR-SWI \citep{2018baupau} using the surface model SURFEX (Surface Externalis\'ee; \citealt{2013masle}) and (2)  employing a TCA-based local error approach. The goal is to obtain a level 4 soil moisture product\footnote{A level 4 product is obtained by applying a model or an analysis to a lower level data product.} and to demonstrate its value for the NWP forecast model AROME \citep{2011seibro} in Austria. 

As in situ soil moisture measurements are only sparse in Austria and the very different spatial representativity makes a sensible comparison with remotely sensed or modelled soil moisture difficult, we employ two different approaches for the verification of the soil moisture analysis. First, we exploit the high spatial and temporal coverage of the Austrian semi-automatic weather stations. We use their 2~m temperature and 2~m relative humidity measurements as reference for the forecasted 2~m values of our assimilation system and, therefore, as indirect measure for the performance of the soil moisture assimilation. Second, we employ a gridded water-balance product as a direct reference when comparing the soil moisture analyses of our assimilation experiments. The water balance has been used in numerous studies as a proxy for soil moisture \citep{2012muesen,2016herkal,2019hashof}.

This manuscript is organised as follows. Section~\ref{sec:dat} will present the observations that are used for data assimilation, as well as the additional data sets needed for bias correction, for the estimation of the observation errors, and for the verification. The SURFEX Offline Data Assimilation and the employed Kalman Filter will be introduced in Sect.~\ref{sec:soda}.
The characterisation of the observation errors with TCA will be described in Sect.~\ref{sec:tca}. Its implementation and the general setup of the data assimilation system will be given in Sect.~\ref{sec:sda}.
The performance of the data assimilation and the verification of the results will be presented and discussed in Sect.~\ref{sec:res}. The results will be summarised in Sect.~\ref{sec:sum}.

\section{Data and preprocessing} \label{sec:dat}
\subsection{SCATSAR-SWI} \label{sec:swi}
The SCATSAR-SWI \citep{2018baupau} combines 25~km MetOp/ASCAT and 1~km \mbox{Sentinel-1/CSAR} surface-soil moisture retrievals and aims at exploiting both the high temporal resolution of the former and the high spatial resolution of the latter. In addition, the SCATSAR-SWI provides vertically resolved soil moisture information by taking into account the propagation of the measured surface soil moisture into deeper soil layers through applying an exponential filter approach \citep{1999waglema}. The soil layers are quantified by the $T$-parameter, where $T$ represents the infiltration time in days. The SCATSAR-SWI covers most of Europe and is available in near-real time from the Copernicus Global Land Service\footnote{\url{https://land.copernicus.vgt.vito.be/PDF/portal/Application.html}} including dates back to 2015.

The data employed for our study are obtained directly by the TU Wien with a 500~m grid sampling, which corresponds to a spatial resolution of 1~km. They are delivered with the Equi7Grid \citep{2014bausab} as a spatial reference system. 
For covering the study area (the Austrian domain), two neighbouring, but not overlapping data tiles (each 600~km x 600~km) are merged into one file. The obtained data set is upscaled to the 2.5~km grid of the assimilation system using linear interpolation, which reproduced the original data simultaneously the best and the most efficiently of all tested methods. The units are converted from percentaged soil-water index (SWI) to fractional volumetric soil moisture (WG) using the following relationship:
\begin{equation}
    \mathrm{SWI} = \frac{\mathrm{WG} - \mathrm{WG}_\mathrm{min}}
    {\mathrm{WG}_\mathrm{max} - \mathrm{WG}_\mathrm{min}} =
    \frac{\mathrm{WG}}{w_\mathrm{sat}} \;, \label{eq:swi}
\end{equation}
where we assume $\mathrm{WG}_\mathrm{min}=0$ and $\mathrm{WG}_\mathrm{max}=w_\mathrm{sat}$. The saturation values $w_\mathrm{sat}$ are derived from the sand fraction using the SURFEX model \citep{2011decboo}.

We utilise the six uppermost $T$-values ($T=2,5,10,20,40,60$~d) for data assimilation. The SCATSAR-SWI of these values are reasonably well correlated to the six uppermost layers of the soil model (Pearson correlation coefficient $r_P>0.3$) and provide continuous input for those soil layers (depth~$=1,4,10,20,60$~cm) in the data assimilation. The soil moisture variables of these layers will be denoted WG1 to WG6 in the following.

For bias correction (Sect.~\ref{sec:dat}\ref{sec:bc}), data from $2015 - 2017$ are used as reference. For TCA (Sect.~\ref{sec:tca}) and for data assimilation (Sect.~\ref{sec:sda}), the computations are performed on data from 2018.

\subsection{SMAP}
To fulfil the requirements for TCA (Sect.~\ref{sec:tca}), two independent data sets are required in addition to the SCATSAR-SWI observations. As those are actively sensed, one of the additional data sets can be a soil moisture data set of a passive sensor. The radiometer onboard the Soil Moisture Active Passive (SMAP) mission promised the best data quality among several considered passive microwave instruments.
Here, the downscaled product ``SMAP Enhanced L3 Radiometer Global Daily 9 km EASE-Grid Soil Moisture, Version 2'' is used \citep{2018oncha}. Ascending and descending data are merged into one data set per day.
Analogous to the SCATSAR-SWI, the SMAP data (volumetric soil moisture) is linearly interpolated to the 2.5~km grid of the model.

\subsection{Water balance} \label{sec:datwb}
In a simplistic approach, when runoff is negligible, the (climatic) water balance describes the difference between water supply and water demand.
Using daily gridded precipitation \citep{2018hiefre} and potential evapotranspiration \citep{2016hasbar} fields over Austria, we compute the water balance $\mathrm{WB}$ by subtracting the potential evapotranspiration $ET$ from the precipitation $P$:
\begin{equation}
    \mathrm{WB} \approx P - ET \;.
\end{equation}
The data sets have a spatial resolution of 1~km and cover not only Austria itself but include also close-by catchments. Uncertainties in assessing the water balance arise from those inherent from the input data sets. The gridded precipitation data is based on a fixed station network from 1961 onward, which enables climatological consistency over time on one hand, but may lack spatial consistency due to a limited number of gauges used for interpolation. The potential evapotranspiration data set is calculated using a simple temperature based approach \citep{1975har,2003harall} but applying a re-calibration procedure to minimise the bias compared to the FAO\footnote{Food and Agriculture Organization of the United Nations} recommended Penman-Monteith formulation \citep{1998allper}. See \citet{2016hasbar} for further details. 

To be able to evaluate the soil moisture of all assimilated soil depths, the water balance is processed in four steps, which will be described in more detail:
\begin{enumerate}\itemsep=-2pt
    \item Interpolation to the model grid.
    \item Creation of a multi-layer product.
    \item Conversion to an SWI-like measure.
    \item Minimisation of the bias.
\end{enumerate}

Step 1: The water balance is interpolated to the 2.5~km grid of the soil model using the nearest neighbour method. This method reproduced the original data better than linear interpolation in this case due to the irregular border of the data.

Step 2: To derive a multi-layer product, we apply a method similar to the exponential filter described in Sect.~\ref{sec:dat}\ref{sec:swi}. Hereby, the water balance $\mathrm{WB}(t)$, with time $t$ in days here, is smoothed using a weighted average over the last $n$ days :
\begin{equation}
    \mathrm{WB}_{t_n} = \frac{\sum_{i=1}^{n} w(t_i) \mathrm{WB}(t_i) }{\sum_{i=1}^{n} w(t_i)}
\end{equation}
In a first approach, a linear weighting function $w(t)$ was constructed such that the weights $w(t_0)=1$ and $w(t_n)=0$. Further tests revealed, however, that an exponential function is more suited to reproduce the modelled soil moisture time series, especially for the declining slopes.
Using the exponential weights, the average $\mathrm{WB}_{t_n}$ is computed for time ranges $n={1,...,200}$ days. The time series obtained in that fashion are compared with the model soil moisture at different depths to find the averaging time length that results in the best correlation for each soil layer.

We obtain realistic values for the soil layers 2 to 5. The uppermost and lowest assimilated soil layer cannot be reproduced realistically by the water balance, due to a too large (layer 1) or a too small (layer 6) temporal variability, respectively. We will therefore omit these layers in the evaluation.
We note that the compared products represent information obtained on very different time scales (days for the water balance vs. several minutes for the soil model), which reduces the comparability especially for the uppermost soil layer.

Step 3: For making both soil moisture analysis and water balance comparable, we transform the soil moisture analysis from volumetric soil moisture to SWI (eq.~\ref{eq:swi}).
Similarly, we scale the water balance to values between 0 and 1 using \begin{equation}
    \mathrm{WB}_\mathrm{scaled} = \frac{\mathrm{WB} - \mathrm{WB}_\mathrm{min}}
    {\mathrm{WB}_\mathrm{max} - \mathrm{WB}_\mathrm{min}} \;,
\end{equation}
where $\mathrm{WB}_\mathrm{min}$ and $\mathrm{WB}_\mathrm{max}$ are the domain-wide minimum and maximum values of the water balance.

Step 4: As the water balance computed in this way reveals a large bias compared to the model soil moisture, the global minimum and maximum values are optimised via a brute-force search to yield the smallest possible bias.

\subsection{Bias correction} \label{sec:bc}
Systematic errors in the observations can prevent the data assimilation algorithm from converging. This can be prevented by removing a possible bias beforehand \citep{1998deeda}. The bias correction can be achieved with so-called CDF (Cumulative Distribution Function) matching (e.g., \citealt{2004reikos}). 
Using this method, the CDFs of the SCATSAR-SWI are matched to the CDFs of a reference data set (here, SURFEX model soil moisture) by performing a 4th-order polynomial fit of the time series for every grid point in the domain in a reference time period (here, $2015 - 2017$). Under the assumption that the bias does not change between the epochs, the resulting fit functions are used to remove the bias in the data set of 2018. To achieve this, we partially employ routines of the \texttt{pytesmo} package of the TU Wien (v0.6.8; \citealt{pytesmo068}).

\section{SURFEX Offline Data Assimilation} \label{sec:soda}

For assimilating the SCATSAR-SWI, the SURFEX Offline Data Assimilation (SODA) is employed using SURFEX v8.1 \citep{2013masle} with a simplified extended Kalman filter (sEKF; \citealt{2009mahber}).
Within SURFEX, the soil is described with the ISBA model (Interaction Sol-Biosph\`ere-Atmosph\`ere; \citealt{1989noipla}) using the diffusion scheme ISBA-DIF \citep{2011decboo}, which can treat up to 14 soil layers down to 12~m.
We adapted the SODA code to be able to use six soil layers for data assimilation and read the local observation errors from a file.

The atmospheric forcing needed for SODA is provided by short-range forecasts (up to +24~hours) of the NWP model AROME (version cy40t1). In return, AROME is initialised with a surface file containing the information of the soil moisture analysis at 12~UTC using a 24~h assimilation window. The forecasted 2~m temperature and 2~m relative humidity of AROME will be used for verification. This setup describes a so-called quasi weakly-coupled data assimilation system as defined in \citet{2009fujnak} and \citet{2017penake}.

In the Kalman filter \citep{1960kal}, the prediction of the state vector $\bm x_\mathrm{b}$ (background) is updated with the observations $\bm y$ to the analysis
\begin{equation}
	\bm x_\mathrm{a} = \bm x_\mathrm{b} + \bm d^\mathrm{a}_\mathrm{b} \label{eq:kf}
\end{equation}
by adding the analysis increment 
\begin{equation}
	\bm d^\mathrm{a}_\mathrm{b} = \pmb K \bm d^\mathrm{o}_\mathrm{b} \;. \label{eq:ic}
\end{equation}
For the increment, the innovation 
\begin{equation}
	\bm d^\mathrm{o}_\mathrm{b} = \bm y - \pmb H \bm x_\mathrm{b} \label{eq:iv}
\end{equation}
is weighted with the Kalman gain
\begin{equation}
\pmb K = \pmb B \pmb H^T (\pmb H \pmb B \pmb H^T + \pmb R)^{-1}  \;.  \label{eq:kg}
\end{equation}
Here, $\pmb H$ denotes the linearised observation operator (e.g., \citealt{2009mahber}). Its elements $H_{jk}$ can be estimated numerically with finite differences \citep{2004balbou}:
\begin{equation}
H_{jk} = \frac{\partial y_j}{\partial x_k} = \frac{y_j(\bm x+\delta x_k) - y_j(\bm x)}{\delta x_k} \;. \label{eq:jac}
\end{equation}
Here, $y_j(\bm x+\delta x_k)$ and $y_j(\bm x)$ describe the perturbed and the unperturbed model trajectory, respectively, where the perturbation amplitude $\delta x_k=10^{-3} (w_\mathrm{fc}-w_\mathrm{wilt})$ with the field capacity $w_\mathrm{fc}$ and the wilting point $w_\mathrm{wilt}$ for $j,k \in \{1,...,6\}$ in our case. To distinguish between soil moisture observation and control variables in the following, we denote the former with WG$_j$ and the latter with wg$_k$.
$\pmb B$ describes the background error covariance matrix, which is not propagated in time in the sEKF, and $\pmb R$ the error covariance matrix of the observations. 
The validity of the linear assumption for the observation operator can depend on the length of the assimilation window. While a separate study on the optimal length could be worthwhile in principle, the 24~h window is often used in similar studies \citep{2010mah,2014barcal,2017faibar}. Similar to \citet{2009mahber} and \citet{2010rudalb}, we compared the Jacobians obtained with positive and negative perturbations, respectively, and found that the vast majority of points fulfil the assumption of linearity here.

In the next section, we describe the determination of the diagonal entries of $\pmb R$, assuming that the errors of different observations are not correlated.
Regarding the retrieval method of our observations, where the information for the profile soil layers is derived with the exponential filter from the surface soil moisture (Sect.~\ref{sec:dat}\ref{sec:swi}), this is certainly not the case in reality. The determination of the off-diagonal entries is possible in principle using a fourth data set \citep{2016grusua} but acquiring a fourth independent data set covering the same domain is not feasible here.

\section{Triple Collocation Analysis} \label{sec:tca}

Random observation errors can be characterised using Triple Collocation Analysis \citep{1993ros,1998sto}. The method assumes a linear relation between a measurement $\Theta_\mathrm{m}$ and the true state $\Theta_\mathrm{t}$ of the measured variable:
\begin{equation}
    \mathrm{\Theta}_\mathrm{m} = \alpha + \beta \Theta_\mathrm{t} + \epsilon\;, 
\end{equation}
where $\epsilon$ represents the random error, $\alpha$ the systematic additive bias, and $\beta$ the systematic multiplicative bias.
The resulting equation system can be solved analytically for the error variances $\sigma_{\epsilon}^2$ of a triplet of data sets under a number of assumptions \citep{2014yilcro,2016grusu}. 
The error variances obtained this way represent the diagonal elements of the error covariance matrix of the observations $R_{jj} = \sigma_{\epsilon_j}^2$.

One of the assumptions for TCA is the requirement of zero error cross-correlation between the data sets. This is satisfied by data sets that are as independently obtained and processed as possible. One possible triplet that is commonly assumed to fulfil this requirement is a combination of one active-microwave based, one passive-microwave based, and one modelled soil moisture data set. As a consequence, a data set obtained by a passive satellite sensor (SMAP; Sect.~\ref{sec:dat}) and a data set created with the surface model SURFEX are used in addition to the actively sensed SCATSAR-SWI.

\begin{figure*}[h!]
	\centering
		\includegraphics[width=\textwidth]{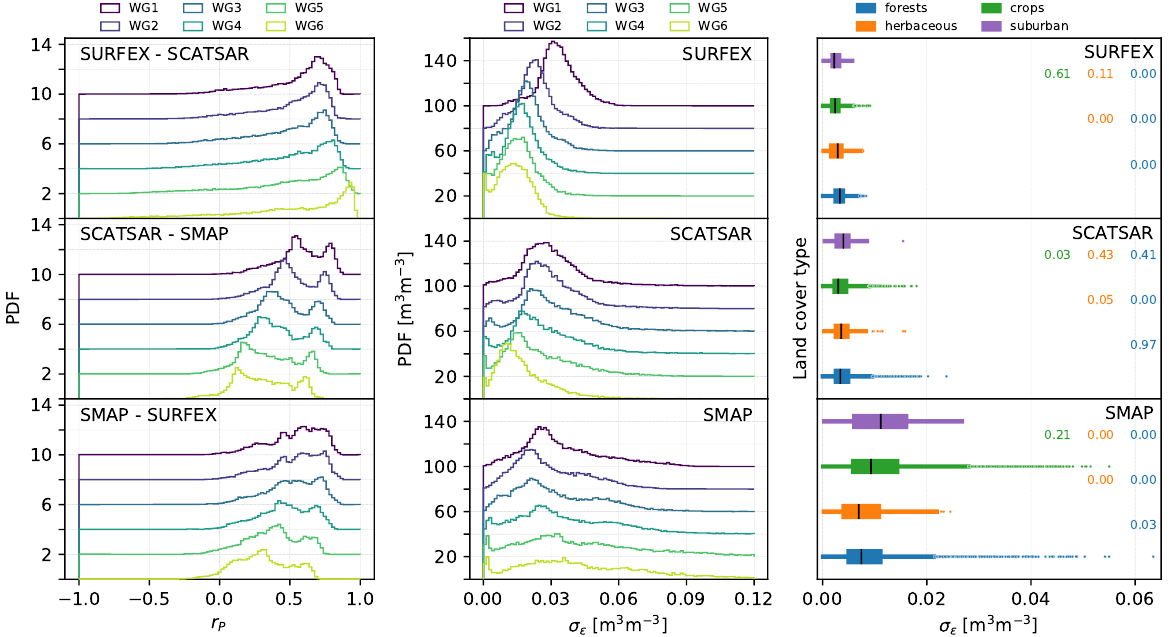}
    \caption{Probability density functions (PDF) of the correlation coefficient $r_P$ (\textit{left}) and of the error STD $\sigma_\epsilon$ obtained with TCA (\textit{centre}) for the three employed data sets and for the soil layers WG1 to WG6. 
    The histograms were stacked using bottom lines separated by a value of 2 for $r_P$ and 20 for $\sigma_\epsilon$.
    \textit{Right}: Box plots of the layer-averaged observation error STD of the three data sets. Each land cover type is displayed separately. The median is indicated with a black line and the $p$-values are given next to each distribution for all combinations with the text colour corresponding to the cover type that was tested against.
    } \label{fig:tch}
\end{figure*}

The error variances and the resulting error standard deviation (STD) of the three data sets are computed for 2018 using the Triple Collocation routine of the \texttt{pytesmo} package of the TU Wien (v0.7.1; \citealt{pytesmo71}). The method is applied to each grid point in the considered domain. Only days where all three data sets have measurements are taken into account since the temporal gaps between two measurements are often too large for realistic interpolation.
In addition, the Pearson correlation coefficient $r_P$ and the associated significance $p$ of each pair of data sets are computed for each grid point. The points without significant positive correlation ($p<0.05$) between all pairs are discarded for the TCA (cf. \citealt{2008scihol}). 

As expected, the single-layer SMAP soil moisture correlates less with the multi-layer data sets with deeper soil layers (Fig.~\ref{fig:tch}, left).
The average correlation between the SCATSAR-SWI and the SURFEX data set improves with deeper soil layers. On the one hand, this can be attributed to the decreasing influence of errors in the atmospheric forcing onto the deeper soil layers in SURFEX. 
On the other hand, deeper soil moisture layers of the SCATSAR-SWI are smoothed by the exponential filter (cf. Sect.~\ref{sec:dat}\ref{sec:swi}), thereby reducing the non-systematic retrieval errors.
This behaviour is mainly seen in the flatlands, whereas the correlation degrades for mountainous terrain where both SURFEX and SCATSAR-SWI are prone to errors related to the topography.

\begin{figure*}[h!]
	\centering
	\includegraphics[width=\textwidth]{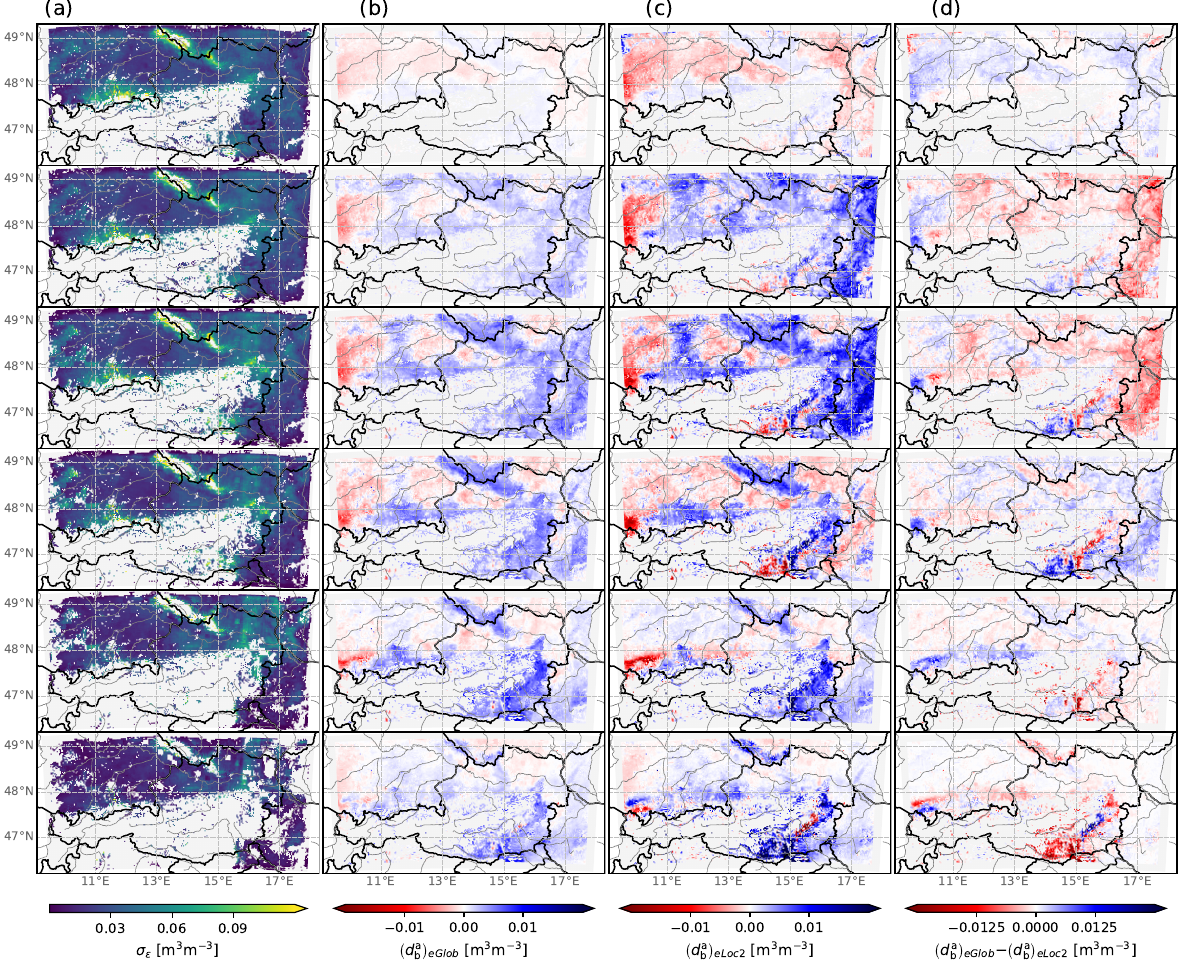}
    \caption{(\textit{a}) Spatial distribution of the observation error STD (see Sect.~\ref{sec:tca}). Averaged analysis increments $\bm d^\mathrm{a}_\mathrm{b}$: (\textit{b}) for \textit{eGlob}, (\textit{c}) for \textit{eLoc2}, and (\textit{d}) for their difference (see Sect.~\ref{sec:res}\ref{sec:astat}). Within each panel, the respective quantity is shown for all soil layers WG1 to WG6 from top to bottom.
    } \label{fig:maps}
\end{figure*}

The three data sets are mutually not correlated significantly for many points for deeper soil layers. This results in a decrease of the number of grid points where it was possible to compute the error variances/STD with TCA (Fig.~\ref{fig:maps}a). 
In those parts of mountainous regions that were not already masked in the input data, the error STD is significantly higher than in the flatlands, which reflects the known physical limits of active satellite sensors in forested areas and complex terrain \citep{2012drarei,2019baufre}. We note that the exponential filter is applied to all pixels in the same way, not distinguishing between type of terrain. However, any masking of input surface soil moisture is forwarded to the SWI calculation, neglecting affected surface soil moisture values, and decreasing the value of the employed quality flag.

The average error STD of the SCATSAR-SWI decreases with increasing soil-layer depth (Fig.~\ref{fig:tch}, centre). Analogous to the processes describing the behaviour of the correlation coefficients, this is a result of the exponential filter and the resulting smoothing of the soil moisture signal in deeper soil layers. In addition, the variability in the deeper layers is naturally lower, which is reproduced as well by the soil model. Furthermore, the SCATSAR-SWI error distribution is more similar to the SMAP error distribution for the upper layer and to the SURFEX error distribution for the lower layers, which reflects the matching variability characteristics of the different data sets.

The magnitude of the obtained error STD ($\sigma_\epsilon \approx 0.025$~m$^3$m$^{-3}$ on average) is comparable with the values found in other studies for similar soil moisture data sets: $0.05\pm0.05$~m$^3$m$^{-3}$  for SMMR \citep{2008dewag} and 0.02 to 0.07~m$^3$m$^{-3}$ for ASCAT and AMSR-E \citep{2010dorsci,2011dramah,2011loesch,2013drarei}.

Due to different performances of the backscatter retrieval algorithm over different land cover types \citep{1999waglem} and the deteriorating effect of dense vegetation on radiometer data \citep{2010dorsci}, a dependency of the TCA-error distributions on land cover type seems likely.
We extract the land cover types from ECOCLIMAP physiographic data \citep{2003mascha} and test for a dependency of the error STD on the land cover type, also with regard to the topographic complexity (cf. \citealt{2012drarei}). We assign a grid point to a certain cover type if its cover fraction is larger than 0.5 and apply a Mann-Whitney-U test \citep{1947manwhi} to test if the error distributions differ significantly from each other. 

We found that the majority of all tested combinations of the land-cover separated error distributions of the three data sets differ significantly from each other (Fig.~\ref{fig:tch}, right). Masking the high ($>10\%$) topographic complexities only results in minor statistical changes compared to the original distributions since most of the topographically complex areas are already masked in the observations.

The agreement of the distributions of the different cover types is not consistent for the three data sets, e.g. for the SCATSAR-SWI, only the distributions for crops and forest as well as crops and suburban show a significant difference between each other.
For the SMAP data, only the error distributions of points assigned with suburban and crops cover seem to be part of the same distribution, for SURFEX, additionally the distributions of suburban and herbaceous cover types.
These findings show that the different land-cover dependencies of the three data sets are propagated to the error STD itself but, especially for the SCATSAR-SWI, do not leave a strong signature. 

Eventually, we compare the number of points, where the computation of the TCA error of the SCATSAR-SWI was not successful, with the total number of grid points to be assimilated. We find that the lost points represent fractions of 9, 10, 12, 14, 23, and 35\% for soil layer 1 to 6, respectively. The cover types forest, herbaceous and suburban exhibit similar fail rates of between 15 and 46\% for the single soil layers. As expected, the smallest loss is found for points with crop cover ($\approx 1 - 27$\%). This is the cover type the most present in less complex terrain where all three data sets can be assumed to have the largest reliability.

\section{Setup of the data assimilation system} \label{sec:sda}

For implementing the observation error STD obtained with TCA into SODA, the STD values are converted from fractional volumetric soil moisture to fractional SWI (Sect.~\ref{sec:dat}\ref{sec:swi}, eq.~\ref{eq:swi}). On grid points where the TCA does not yield a value, the average value of the error STD of the respective soil layer is utilised instead. 

Due to the high amount of solar radiation in summer and the resulting higher rates of water and energy exchanges, the impact of soil moisture on the atmosphere is expected to be largest in the summer months. In addition, the spatial coverage of the SCATSAR-SWI is best in the warm season when the soil is mostly unfrozen. As a consequence, the months April to September 2018 are chosen as testing period for data assimilation. As the data assimilation system needs around a fortnight to spin up, the data assimilation system is run for the month before the actual period of interest (i.e. for March 2018). 

As a reference experiment, SODA is run with a global error ($\sigma_\epsilon=0.2$ in units of SWI), referred to as \textit{eGlob}. Two different approaches for the local TCA-based observation error estimates are tested, denoted \textit{eLoc1} and \textit{eLoc2}, respectively, and described as follows. The background error remains unchanged ($\sigma_B=0.2$) for all experiments.

\textit{eLoc1}: To assess the impact of the spatial variability of the observation errors, the obtained error STD are scaled by a constant factor so that their spatial average matches the global observation error. The error values of the uppermost layer are on average smaller than the global observation error by a factor of $\approx 3$. By scaling with this value, the average error STD of the uppermost soil layer is kept at the value of the global observation error. To preserve the vertical error pattern, the average errors of the other five layers are adjusted by this to 0.19, 0.20, 0.18, 0.16, and 0.11, respectively.
In addition, this method prevents the Kalman gain from approaching unity for the smallest observation errors, which would neglect the influence of the model background in the resulting analysis, especially for the deeper layers.

\textit{eLoc2}: Using the TCA error distribution as it is will reveal to what extent the supposedly more realistic absolute magnitude of the observation errors can improve the analysis.
As mentioned, the obtained error values are on average several times smaller than the global observation error, which will cause the assimilation to put more emphasis on the observations than on the model background compared to \textit{eGlob}. 

\section{Results and discussion} \label{sec:res}

In the following, the results of the computations using global and local observation errors as described in the previous section will be set in contrast to each other. More specifically, the performance of the data assimilation system will be evaluated in general, the atmospheric forecast and the soil moisture analysis will be verified individually. 

\subsection{Evaluation of the Jacobians} 
One critical aspect of data assimilation using the sEKF is the computation of the Jacobian matrix of the observation operator $H_{jk}$ (eq.~\ref{eq:jac}), which is needed for the transformation between observation ($y_j$) and model ($x_k$) space. 
Here, the observation vector consists of the six soil layers of the SCATSAR-SWI (WG1 $-$ WG6) described in Sect.~\ref{sec:dat}\ref{sec:swi}, and the control vector of the six upper soil layers of the SURFEX model (wg1 $-$ wg6).

\begin{figure*}[htb]
	\centering
	\includegraphics[width=\textwidth]{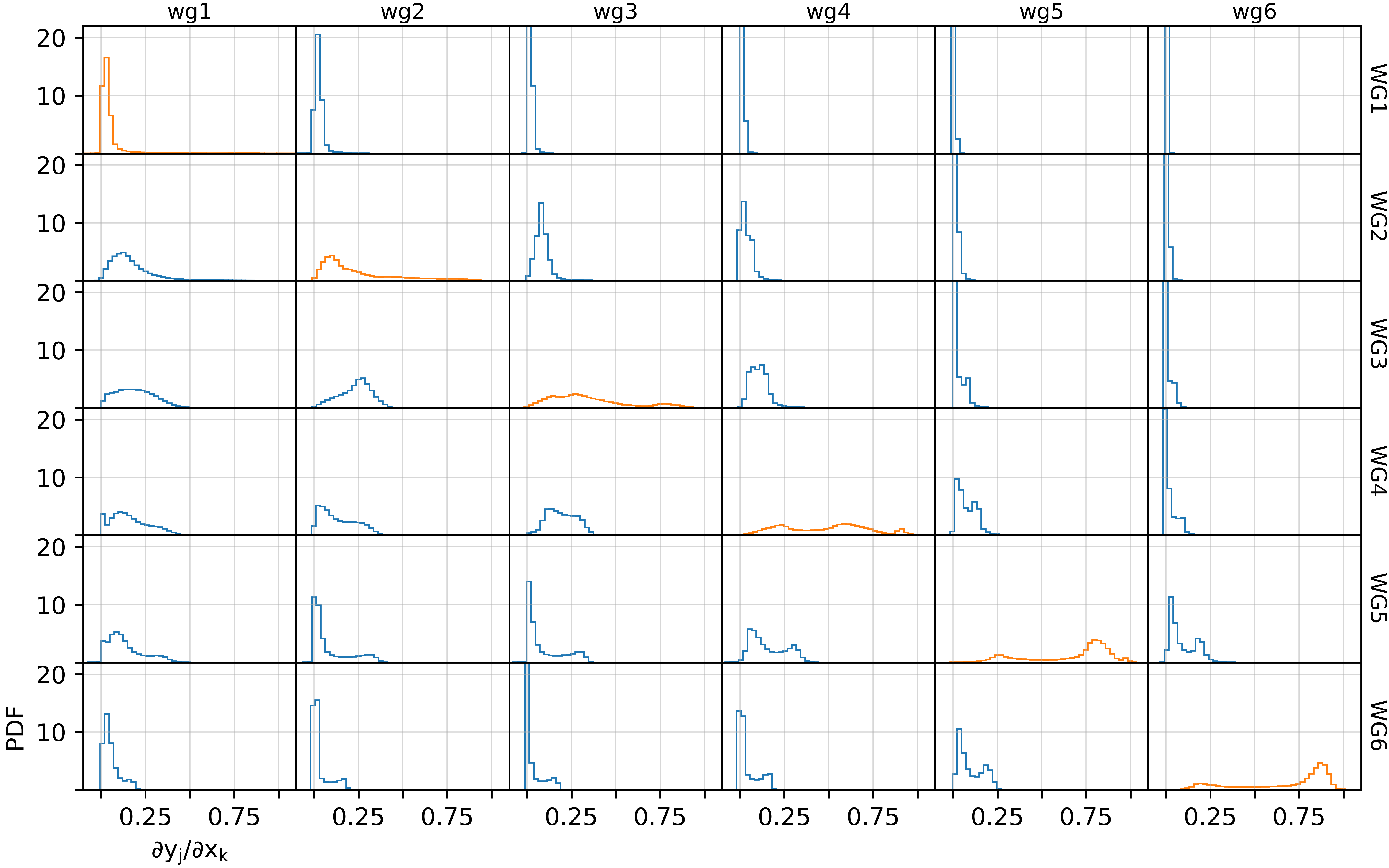}
	\caption{Probability density functions of Jacobians $\frac{\partial y_j}{\partial x_k}$ for observation variables $y_j$ (WG1~$-$~WG6) and control variables $x_k$ (wg1~$-$~wg6). The Jacobians were accumulated over the whole time range considered for data assimilation. Bins with PDF-values larger than 22 are cropped at that value for better visualisation. 
    } \label{fig:jacsh}
\end{figure*}

The values of the Jacobians estimate the sensitivity of an observation variable with respect to a specific control variable. 
Similar to other studies \citep{2010rudalb,2017albmun}, the Jacobians obtained here adopt approximately three different distributions: Characteristic are the peak at zero, the peak at 1, and rather dispersed distributions. These distribution types are visible in the distributions that are averaged over the whole investigated time range (Fig.~\ref{fig:jacsh}) as well as in the monthly statistics (not shown). 

\begin{sidewaysfigure*}[p]
	\centering
	\includegraphics[width=\textwidth]{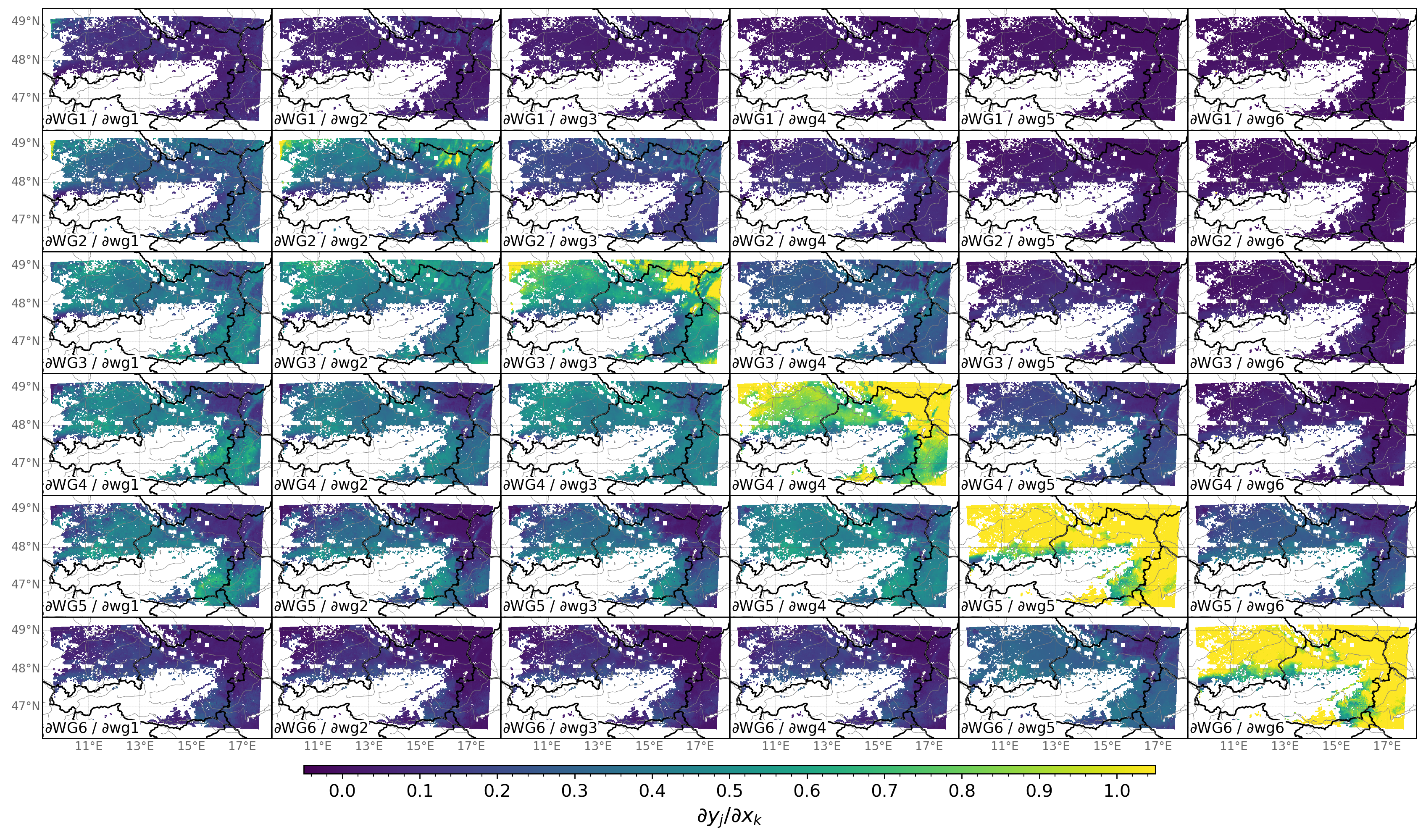}
	\caption{Spatial distributions of temporally averaged Jacobians $\frac{\partial y_j}{\partial x_k}$ for observation variables $y_j$ (WG1~$-$~WG6) and control variables $x_k$ (wg1~$-$~wg6). 
    } \label{fig:jacsm}
\end{sidewaysfigure*}

Close-to-zero Jacobians indicate that the sensitivity to a specific control variable is low. This is mainly the case for the lowest and uppermost soil-layer observation variables, respectively. The upper soil layers are almost insensitive to perturbations of wg6. WG1 is barely sensitive to perturbations of any control variable due to the different time scales the variables are representative for. 

For an observation variable at a given soil layer, the sensitivity to the upper soil layer control variables is higher than to the lower ones. This might be due to the dominance of top-down processes in the model such as infiltration over bottom-up processes such as capillarity (cf. \citealt{2009dramah}), possibly due to different characteristic time scales.
With increasing soil layer depth, the diagonal Jacobians $H_{ii}$ shift more towards one, indicating a more and more linear dependence.

Different sensitivities in flatlands and mildly mountainous regions (Fig.~\ref{fig:jacsm}) sometimes cause bimodal distributions in the histograms, especially between May and August, with the flatlands rather close to zero and the mountainous regions rather larger than 0.5. This indicates that the interaction between the soil layers becomes less in the flatlands with progressing summer. This could be a consequence of soil moisture values approaching either field capacity or the wilting point, where the dynamics of the soil model might be lower (cf. \citealt{2011barcal}). 
Differences between the different assimilation experiments are visible but do not cause changes in the distribution type.

\subsection{Assimilation statistics} \label{sec:astat}

\begin{table*}[htb]
	\centering
	\caption{Mean ($\langle d^\mathrm{o}_\mathrm{b}\rangle$, $\langle d^\mathrm{o}_\mathrm{a}\rangle$) and STD ($\sigma(d^\mathrm{o}_\mathrm{b})$, $\sigma(d^\mathrm{o}_\mathrm{a})$) of innovations and residuals in $10^{-3}$~m$^3$m$^{-3}$, specified and diagnosed error variances ($(\sigma^\mathrm{o}_\mathrm{spec})^2$ and $(\sigma^\mathrm{o}_\mathrm{diag})^2$ after \citealt{2005desber}; see Sect.~\ref{sec:res}\ref{sec:astat}) in $10^{-4}$~m$^3$m$^{-3}$, and ratio $q$ between them for all assimilated soil layers. Each line-separated column contains the values for \textit{eGlob}, \textit{eLoc1}, and \textit{eLoc2} from left to right. Zero values were omitted for the computations.
    } \label{tab:astat}
	\setlength{\tabcolsep}{4pt}
	\begin{tabular}{c|rrr|rrr|rrr|rrr|rrr|rrr|rrr}
	\hline \hline	
	WG & $\langle d^\mathrm{o}_\mathrm{b}\rangle$ &&& $\langle d^\mathrm{o}_\mathrm{a}\rangle$ &&& $\sigma(d^\mathrm{o}_\mathrm{b})$ \hspace*{-5mm} &&& $\sigma(d^\mathrm{o}_\mathrm{a})$ \hspace*{-5mm} &&& $(\sigma^\mathrm{o}_\mathrm{spec})^2$ \hspace*{-6mm} &&& $(\sigma^\mathrm{o}_\mathrm{diag})^2$ \hspace*{-5mm} &&& $q$ &\\
	\hline
1 &   5.9 &   6.3 &   4.5 &   6.1 &   6.6 &   5.2 &   8.1 &   8.7 &   7.4 &   7.8 &   8.3 &   6.8 &   3.1 &   4.2 &  0.51 &  0.99 &   1.1 &  0.72 &   3.1 &   3.7 &  0.71 \\
2 &  0.37 &  0.69 & -0.84 & -0.57 & -0.20 &  -3.0 &   6.4 &   7.0 &   5.8 &   5.5 &   6.3 &   5.5 &   3.1 &   4.0 &  0.49 &  0.34 &  0.43 &  0.28 &   9.0 &   9.4 &   1.8 \\
3 &  -1.0 & -0.66 &  -1.8 &  -2.6 &  -2.2 &  -4.6 &   6.2 &   6.7 &   5.4 &   4.8 &   5.6 &   5.4 &   3.1 &   4.6 &  0.56 &  0.31 &  0.36 &  0.27 &  10.0 &    13 &   2.1 \\
4 &  -1.1 & -0.71 &  -1.6 &  -2.3 &  -1.6 &  -2.1 &   5.3 &   5.6 &   4.6 &   3.6 &   4.1 &   3.7 &   3.1 &   3.8 &  0.47 &  0.20 &  0.22 &  0.12 &    15 &    17 &   3.9 \\
5 &  0.02 &  0.24 & -0.40 &  -1.3 &  -1.0 &  -1.8 &   5.9 &   5.4 &   4.5 &   4.5 &   3.7 &   3.2 &   3.1 &   2.9 &  0.36 &  0.25 &  0.17 &  0.08 &    13 &    17 &   4.2 \\
6 &   1.2 &   1.3 &  0.66 &  0.20 &  0.08 &  -1.0 &   6.4 &   5.5 &   4.5 &   5.2 &   3.8 &   2.7 &   3.1 &   1.4 &  0.17 &  0.32 &  0.20 &  0.02 &   9.5 &   6.9 &   7.2 \\
	\hline
	\end{tabular}
\end{table*}

\begin{figure*}[htb]
	\centering
		\includegraphics[width=\textwidth]{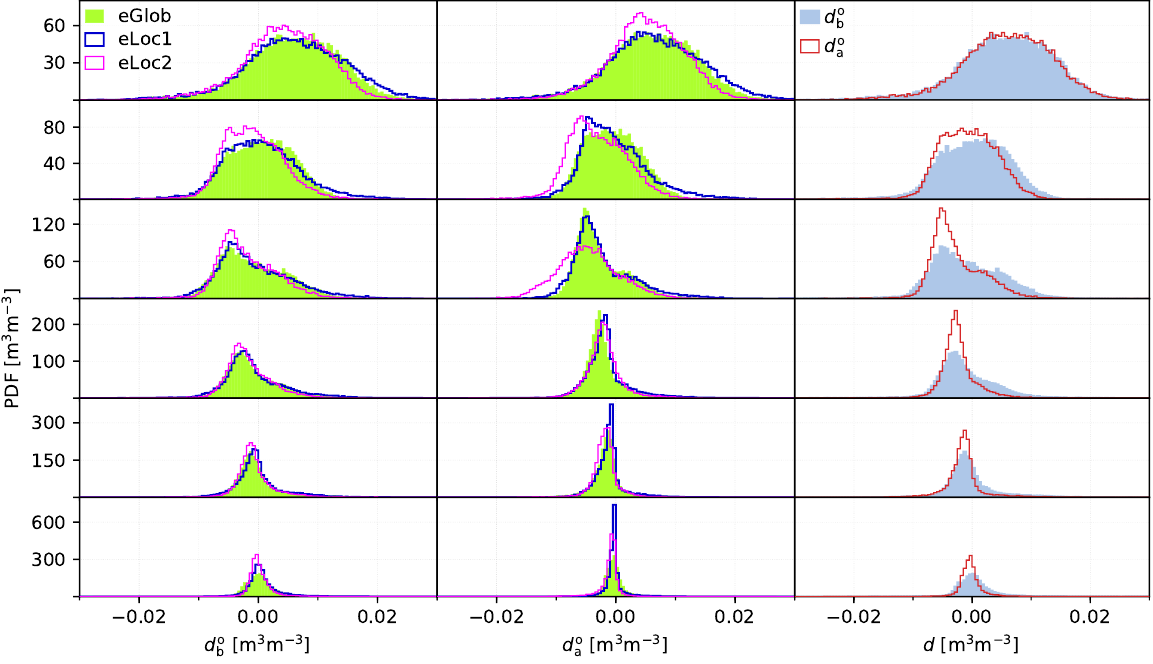}
    \caption{Probability density functions of innovations $d^\mathrm{o}_\mathrm{b}$ (\textit{left}) and residuals $d^\mathrm{o}_\mathrm{a}$ (\textit{centre}) for all experiments. Within each column, each quantity is shown for the soil layers WG1 $-$ WG6 from top to bottom. For \textit{eGlob}, innovations and residuals are shown in direct comparison as well (\textit{right}). Zero values were omitted for the computations.
    } \label{fig:hists}
\end{figure*}

In data assimilation, the analysis is the result of correcting the information from the background state estimate with the observations. Hence, under the assumption of bias-free observations and a statistically relevant number of assimilation cycles, the distribution of analysis residuals $\bm d^\mathrm{o}_\mathrm{a}=\bm y-\pmb H \bm x_\mathrm{a}$ should be more confined around zero compared to the innovations $\bm d^\mathrm{o}_\mathrm{b}=\bm y-\pmb H \bm x_\mathrm{b}$ (Sect.~\ref{sec:soda}, eq.~\ref{eq:iv}) in an effective data assimilation system (e.g., \citealt{2017reide}). 

Concordantly, in our assimilation experiments, the STDs of the residuals are smaller than the STDs of the innovations for all layers in all experiments except WG3 in \textit{eLoc2} (Table~\ref{tab:astat}, Fig.~\ref{fig:hists}). 
Here, the small errors might put a too strong emphasis on the observations in some parts of the domain and cause the experiment to yield a suboptimal analysis. 

The mean innovations and residuals are in general close to zero (-4.6 $-$ 6.6$\cdot 10^{-3}$~m$^3$m$^{-3}$), which illustrates that the systematic errors were corrected adequately (cf. Sect.~\ref{sec:dat}\ref{sec:bc}).
The mean values show that the distributions of the residuals shift more towards negative values compared to the innovations for all experiments and soil layers except WG1.

This behaviour is reflected as well by the dominantly positive increments (Sect.~\ref{sec:soda}, eq.~\ref{eq:ic}; Fig.~\ref{fig:maps}b,c) in all layers except WG1 and WG4, where the increments are more balanced around zero.
Throughout all layers, most of the positive increments are obtained mainly in mountainous regions (Bohemian forest, foothills of Alps), but partially also in Southern Bavaria. For WG2 and WG3, also the flatlands in the East exhibit positive increments. In particular, \textit{eLoc2} exhibits larger corrections in the East of the domain compared to \textit{eGlob}. 
The lower observation quality in regions with complex topography might have led to a suboptimal bias correction and cause the analysis to become too wet here, especially in the deeper soil layers, where errors in the surface soil moisture are added up with the propagation into the lower layers.
The fact that the flatlands exhibit a more mixed behaviour throughout the different layers indicates that, here, the soil moisture is better captured by the observations.

The regions where the differences between the increments of \textit{eGlob} and \textit{eLoc2} are the largest (Fig.~\ref{fig:maps}d) are often regions that obtained a small observation error (Fig.~\ref{fig:maps}a), but the pattern is not consistent. 
The large increments obtained especially for the deeper soil layers in the foothills of the southern Alps originate from assigning the average error STD per soil layer to the empty regions in Fig.~\ref{fig:maps}a. Here, observations and model seem to be badly correlated and the observation error should be larger in principle. The fact that this effect is not captured by the TCA might be a consequence of violated preconditions or indicate that more grid points should have been sorted out on the basis of their significance value (cf. Sect.~\ref{sec:tca}).

The increment maps of \textit{eLoc1} are very similar to the ones of \textit{eGlob}. This shows that a supposedly more realistic error pattern alone does not necessarily lead to large changes on longer time scales.
Furthermore, we do not see any trend in the temporal evolution of the differences between the spatially averaged increments of the experiments. Thus, we do not expect that an assimilation experiment on a longer time period would change the results substantially.
 
The seemingly small impact of the assimilation on the uppermost soil layer could be a consequence of the strong influence of the atmospheric forcing on this layer. 
For example, localised precipitation events, which are difficult to be forecasted correctly, can cause large innovations at single grid cells. The resulting large increments would compensate on average as the probability for both types of wrong forecasts (rain forecasted but not measured and vice versa) is approximately equal in a well calibrated model. 

For further assessing the assimilation performance, we compare specified and diagnosed observation errors after \citet{2005desber}. The Desroziers method requires a linearised observation operator $\pmb H$ as we apply here (Sect.~\ref{sec:soda}). In addition, observation and background errors need to be uncorrelated. Under these assumptions, it is possible to verify whether the errors are specified correctly. In principle, the iterative adaptation of the specified to the diagnosed errors should lead to optimal error values, however, as e.g., in \citet{2011barcal}, we use this method for the diagnosis of our given error distributions only.

Following \citet{2005desber}, the diagnosed observation error can be computed a posteriori for $n$ observations with
\begin{equation}
    \left(\sigma^\mathrm{o}_\mathrm{diag}\right)^2 = \frac{1}{n} \left(\bm d^\mathrm{o}_\mathrm{a}\right)^T \bm d^\mathrm{o}_\mathrm{b} \;.
\end{equation}
This study aims to investigate the impact of different specifications of the observation error and leaves the background error unchanged. Therefore, we focus on the diagnostics of the observation error, similar to other studies \citep{2014wesbel,2016walsim}. Background errors can be assessed in a similar fashion in principle.

The specified error variances for \textit{eGlob} are between 3.1 and 17 times larger than the diagnosed ones (Table~\ref{tab:astat}). The ratios $q=(\sigma^\mathrm{o}_\mathrm{spec})^2/(\sigma^\mathrm{o}_\mathrm{diag})^2$ of \textit{eLoc1} are larger than the ones of \textit{eGlob} except for WG6.
In both experiments, all specified errors are larger than the diagnosed ones, which indicates that the observation error should in principle be smaller. Indeed, \textit{eLoc2} yields smaller ratios, which are closer to the optimal value of one. The ratio for WG6 did not improve compared to \textit{eLoc1} but to \textit{eGlob}. Nevertheless, the results described before suggest that a better agreement of diagnosed and specified errors does not automatically yield a globally improved assimilation performance.

\subsection{Verification of the atmospheric forecast against weather stations} \label{sec:avrf}

To verify the soil moisture analysis indirectly, we evaluate the impact of the different assimilation experiments on the quality of the NWP in Austria by comparing the forecasted 2~m temperature (T2m) and 2~m relative humidity (RH2m) with measurements of the Austrian semi-automated weather stations\footnote{\url{https://www.zamg.ac.at/cms/en/climate/meteorological-network}}. 

The SCATSAR-SWI is not available in mountainous regions owing to the complexity of the radar signal interpretation over strong topography. As no updated analysis is obtained there, only low-elevation stations are taken for comparison. Of all available stations, the ones situated lower than 600~m (around 140) are found to match the area covered by the observations and also by the TCA errors the best.

\begin{figure}[t]
	\centering
	\includegraphics[width=0.49\textwidth]{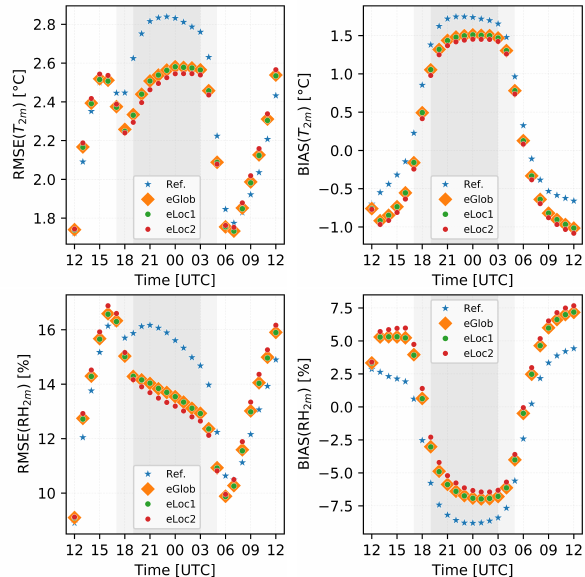}
	\caption{RMSE (\textit{left}) and bias (\textit{right}) of 2~m temperature (\textit{top}) and 2~m relative humidity (\textit{bottom}) for the reference run (blue stars), \textit{eGlob} (orange squares), \textit{eLoc1} (green dots), and \textit{eLoc2} (red dots). 
	The grey shaded areas indicate the approximate duration of the shortest and longest night in Austria in the investigated period. The graphs represent the average over all weather stations below 600~m. 
    } \label{fig:vrf}
\end{figure}

\begin{figure}[h]
	\centering
	\includegraphics[width=0.49\textwidth]{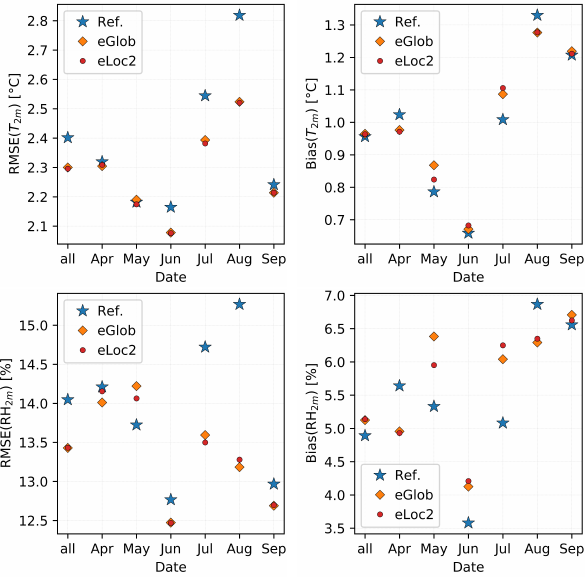}
	\caption{Total and monthly averages of RMSE (\textit{left}) and absolute bias (\textit{right}) of 2~m temperature (\textit{top}) and 2~m relative humidity (\textit{bottom}) for the reference run (blue stars), \textit{eGlob} (orange squares), and \textit{eLoc2} (red dots). The values of \textit{eLoc1} are almost identical with \textit{eGlob} and omitted for clarity.
    } \label{fig:vrfm}
\end{figure}

For the verification, we compute the root mean square error (RMSE) and the bias (forecasted minus measured value) for forecasts up to +24~h with hourly output, averaged over all considered stations (Fig.~\ref{fig:vrf}).
In addition to evaluating the performance metrics of the whole considered time period, each month was investigated separately to detect possible seasonal effects  (Fig.~\ref{fig:vrfm}).
In addition to the assimilation experiments described in Sect.~\ref{sec:sda}, the AROME reference run (without initialisation with the soil moisture analysis) is discussed as well. 

The reference run underestimates the amplitude of the diurnal cycle throughout all investigated months for both T2m and RH2m. This general behaviour is not changed by the assimilation experiments (Fig.~\ref{fig:vrf}, right).

The mostly positive analysis increments (Fig.~\ref{fig:maps}b,c) show that the assimilation adds moisture to the soil, which makes the forecast in the assimilation runs in general colder and more humid than in the reference run. This has a positive impact on the night time forecast, especially for the RMSE, but a negative impact on the day time performance. 
The RMSE of both T2m and RH2m improves for the assimilation runs compared to the reference run for all months except in May.
Furthermore, the RMSE of T2m and RH2m show the largest difference between the reference run and assimilation runs in July and August, where temperature and precipitation have their annual maximum in Austria, hence allowing for strong interactions between the soil and the atmosphere. In September, the differences between the experiments are strongly reduced again -- in agreement with the climatology in Austria and the therefore reduced soil-atmosphere interactions. The trend of the bias is more mixed, with an average advantage of the reference run.

These results demonstrate that the implementation of the soil moisture analysis has the potential to improve the NWP forecasts in spite of the not always optimal assimilation statistics (Sect.~\ref{sec:res}\ref{sec:astat}). This behaviour is likely a consequence of the given adaptation of the AROME configuration to a different treatment of the normally poorly determined soil moisture. The AROME settings for parametrisations, physics, and dynamics themselves remained unchanged for the different experiments.

\begin{figure}[h]
	\centering
		\includegraphics[width=0.5\textwidth]{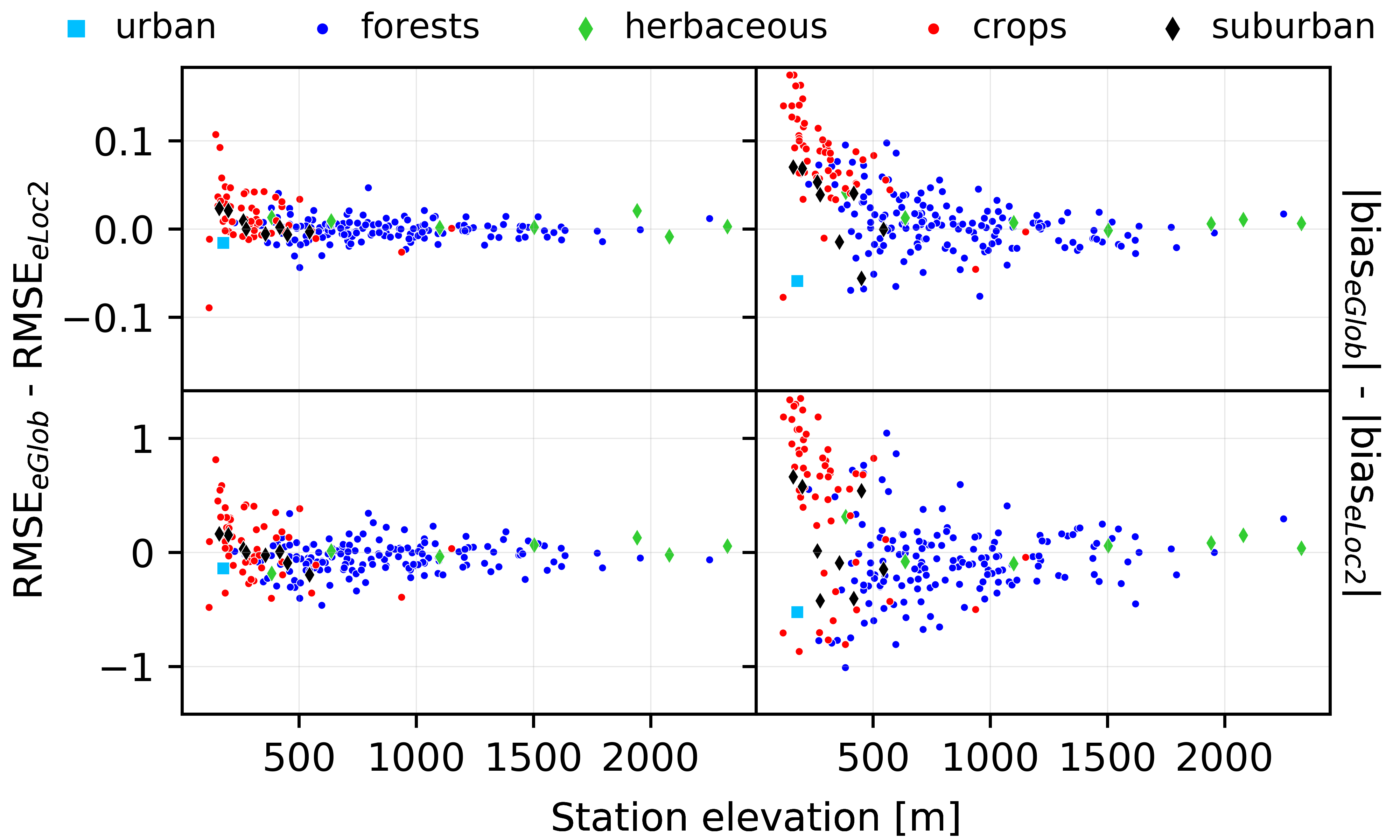}
	\caption{Difference of RMSE (\textit{left}) and absolute bias (\textit{right}) of 2~m temperature (in $^\circ C$, \textit{top}) and 2~m relative humidity (in $\%$, \textit{bottom}) between \textit{eGlob} and \textit{eLoc2} vs. station elevation. Different land cover types are indicated by marker shape and colour. Covers are counted when they have a fraction larger than 0.5. Stations above 2500~m are omitted in the graphs.
    } \label{fig:vrfe}
\end{figure}

Between each other, the assimilation experiments show on average the largest RMSE differences  at night. The experiment \textit{eLoc1} yields mostly identical performance metrics compared to \textit{eGlob}, whereas \textit{eLoc2} shows significant differences for the RMSE and bias of both T2m and RH2m (Fig.~\ref{fig:vrf}). 
With \textit{eLoc2}, the trend of the forecast to become colder and more humid compared to the reference run is sustained. This leads to a degradation of the day time performance and an improvement at night time, averaging to an overall neutral effect of \textit{eLoc2} compared to \textit{eGlob}.

Furthermore, \textit{eLoc2} improves the performance metrics in some months compare to \textit{eGlob}, but shows no consistent behaviour. In general, the differences between the experiments are more pronounced for RH2m than for T2m but do not noticeably differ between each other for both RMSE and bias when averaged over the whole time period.

Nevertheless, single stations can reveal larger differences between the experiments. For this reason, a possible dependence of the difference between the RMSE and bias of \textit{eGlob} and \textit{eLoc2} on the magnitude of the observation error, the station elevation, or the land cover type (cf. Sect.~\ref{sec:tca}) is tested. 
Stations at lower elevations are found to have a larger difference for the RMSE and especially for the bias of both T2m and RH2m (Fig.~\ref{fig:vrfe}). This might be an effect of the more reliable determination of the soil moisture observations in the flatlands.
Subsequently, also the cover types that are predominant at low elevations (e.g., croplands) reveal the largest differences. No dependency on the magnitude of the observation error itself was found.

The results demonstrate that the approach of initialising the atmospheric forecast with a soil moisture analysis employing a local observation error can lead to a slight improvement of the forecast quality of the near-surface prognostic variables dependent on the time of the year. Nevertheless, the actual influence of the local observation error depends strongly on other variables like the station elevation. 
In addition, the results suggest that model-inherent weaknesses in modelling soil moisture are difficult to counteract with performing only one soil moisture analysis per day, especially for the superficial layer.

\subsection{Verification of the soil moisture analysis against the water balance}  \label{sec:wvrf}

\begin{table*}[htb]
	\centering
	\caption{Soil layer, soil layer depth (in cm) and associated averaging time length $t_\mathrm{avg}$ (in days) for the water balance, as well as the mean ubRMSE, bias, and Pearson correlation coefficient $r_P$ between water balance and soil moisture analysis for the three assimilation experiments (ubRMSE and bias in units of fractional SWI). As described in Sect.~\ref{sec:dat}\ref{sec:datwb}, it was not possible to match WG1 and WG6 realistically with the water balance approach.
    } \label{tab:wb}
	\begin{tabular}{cccccc}
	\hline \hline
	WG & depth & $t_\mathrm{avg}$ & ubRMSE            & bias                 & $r_P$   \\ 
	   &       &            & \textit{eGlob} {} \textit{eLoc1} {} \textit{eLoc2} & \textit{eGlob} {} \textit{eLoc1} {} \textit{eLoc2} & \textit{eGlob} {} \textit{eLoc1} {} \textit{eLoc2} \\
   \hline
	2  & 4     & 22               & 0.090 0.090 0.091 & -0.001 -0.001 -0.005 & 0.47 0.47 0.45   \\
	3  & 10    & 26               & 0.084 0.084 0.089 & -0.012 -0.012 -0.016 & 0.47 0.46 0.42   \\
	4  & 20    & 34               & 0.078 0.078 0.084 & -0.017 -0.017 -0.018 & 0.45 0.44 0.39   \\
	5  & 40    & 130              & 0.061 0.062 0.067 & -0.019 -0.020 -0.021 & 0.40 0.39 0.34   \\
	\hline
	\end{tabular}
\end{table*}
\begin{figure*}[htb]
	\centering
	\includegraphics[width=\textwidth]{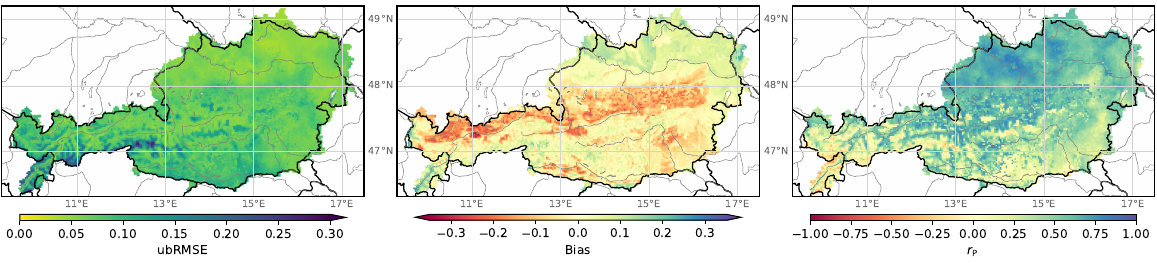}
	\caption{Water balance vs. soil moisture analysis (\textit{\textit{eGlob}}) averaged over all considered soil layers: ubRMSE (\textit{left}), bias (water balance minus soil moisture analysis; \textit{centre}), and correlation (\textit{right}). The ubRMSE and bias are given in units of fractional SWI (see Sect.~\ref{sec:dat}\ref{sec:datwb}).
    } \label{fig:wbmt}
\end{figure*}

For verifying the soil moisture analysis directly, we estimate the soil moisture from a simple water balance model as an additional independent reference (Sect.~\ref{sec:dat}\ref{sec:datwb}).
After the described scaling of the water balance, the average bias (water balance minus soil moisture analysis) is $<0.02$ (in units of fractional SWI; Table~\ref{tab:wb}) and its spatial distribution is such that the soil moisture analysis is wetter than the water balance mostly in the northern Alps and drier in northeast Austria; the behaviour in the southern Alps is more mixed (Fig.~\ref{fig:wbmt}, centre). Due to the artificial tuning of the bias, we will omit a detailed evaluation of this particular metric in the following and focus mostly on the unbiased RMSE (ubRMSE) to minimise the effect of the simplistic bias optimisation.
Following \citet{2010entrei}, the ubRMSE is related to the RMSE and the bias as follows:
\begin{equation}
    \mathrm{ubRMSE}^2 = \mathrm{RMSE}^2 - \mathrm{bias}^2
\end{equation}
and can also be expressed by the correlation $r$ and the standard deviation of the estimated (here, $\sigma_\mathrm{SMA}$ for the soil moisture analysis) and the true measure (here, $\sigma_\mathrm{WB}$ for the water balance):
\begin{equation}
    \mathrm{ubRMSE}^2 = \sigma_\mathrm{SMA}^2 + \sigma_\mathrm{WB}^2 - 2r\sigma_\mathrm{SMA}\sigma_\mathrm{WB} + \mathrm{bias}^2 \;.
\end{equation}

\begin{figure}[t]
	\centering
	\includegraphics[width=0.5\textwidth]{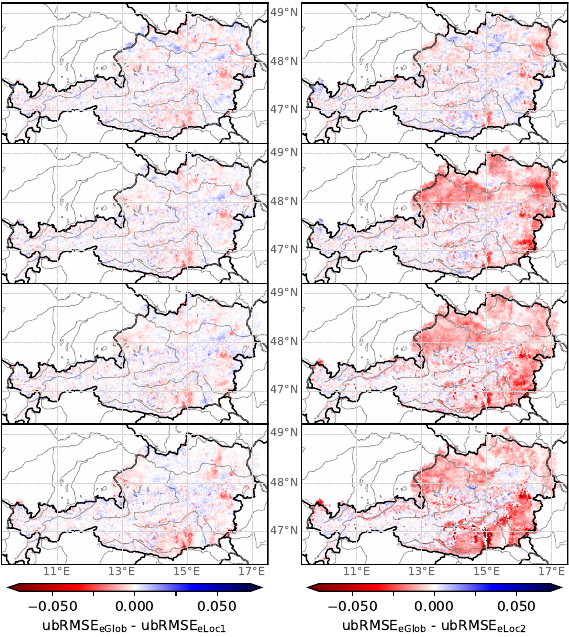}
	\caption{Water balance vs. soil moisture analysis, difference between ubRMSE for WG2 to WG5 (top to bottom): \textit{eGlob} compared to \textit{eLoc1} (\textit{left}) and \textit{eLoc2} (\textit{right}), respectively. 
    } \label{fig:wbdm}
\end{figure}

\begin{figure}[htb]
	\centering
	\includegraphics[width=0.5\textwidth]{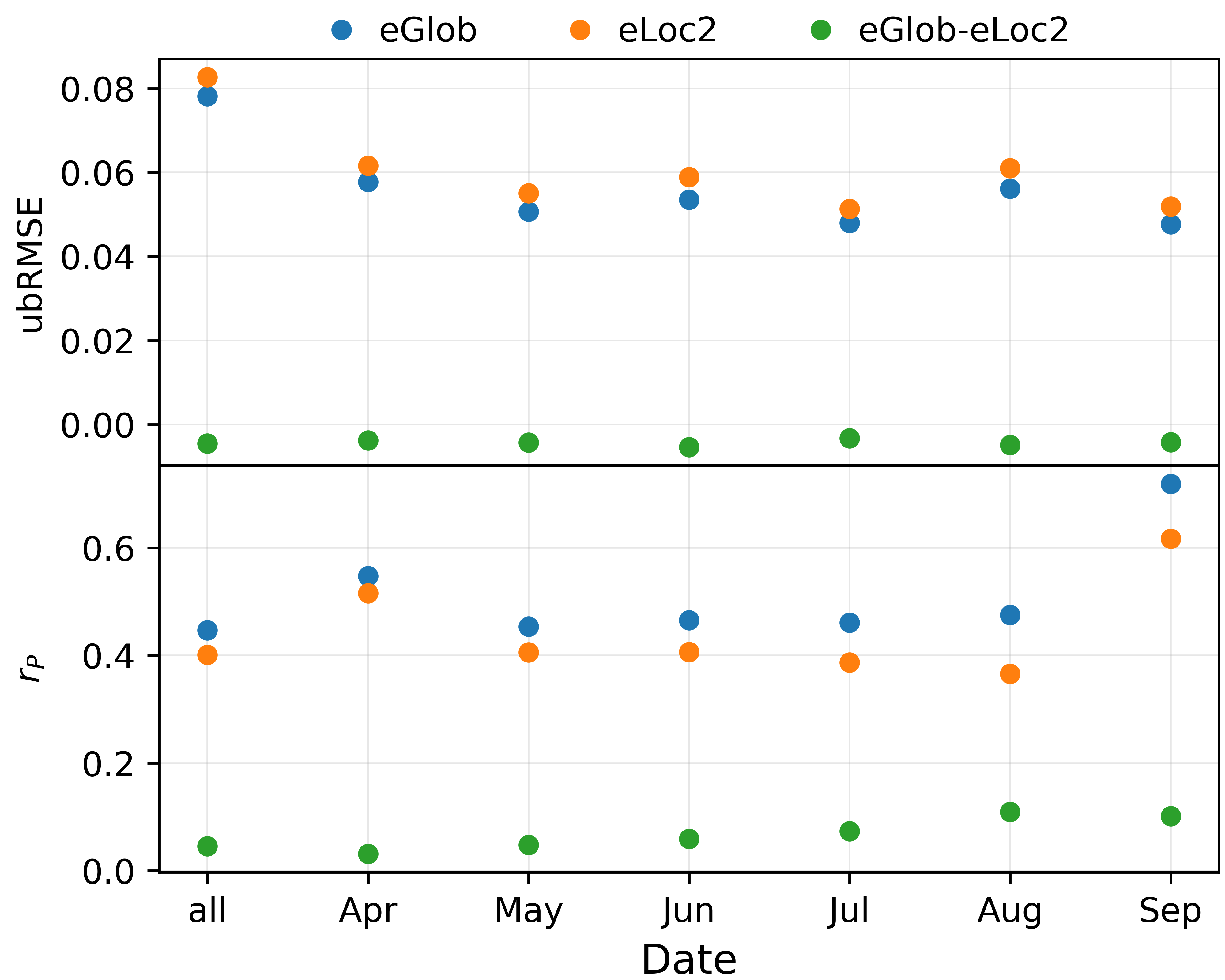}
	\caption{Water balance vs. soil moisture analysis: total and monthly averages of ubRMSE (\textit{top}) and correlation (\textit{bottom}) for \textit{eGlob} and \textit{eLoc2} and their difference. The experiment \textit{eLoc1} is not shown due to its similarity with \textit{eGlob}.
    } \label{fig:wbm}
\end{figure}

\begin{figure}[htb]
	\centering
		\includegraphics[width=0.49\textwidth]{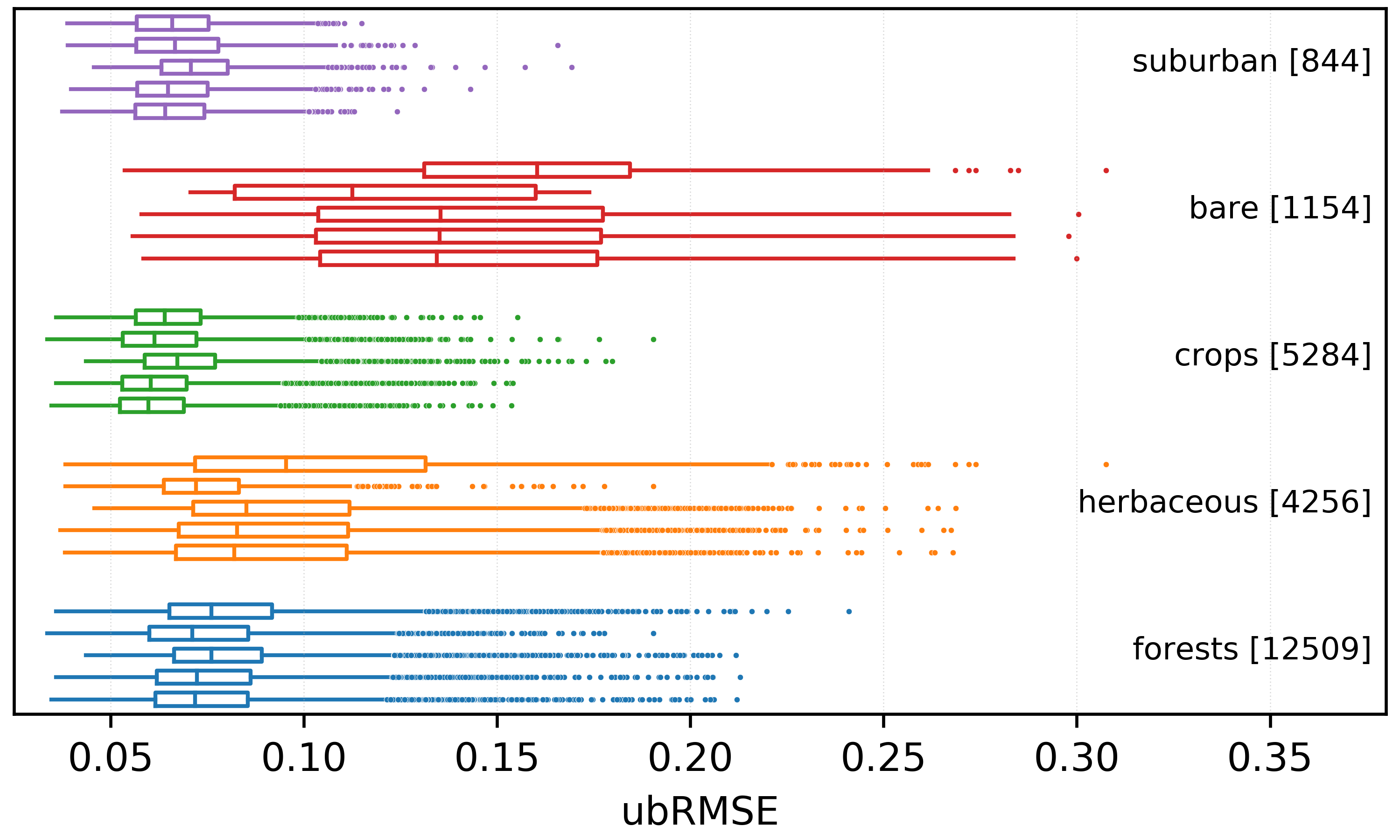}
	\caption{Box plot for the ubRMSE between water balance and soil moisture averaged over all considered soil layers for the different cover types. The number of points with a certain cover type is given in brackets. The topmost box of each colour represents the soil model, the second the observations, the third \textit{eLoc2}, the fourth \textit{eLoc1} and the fifth box \textit{eGlob}.
    } \label{fig:wbb}
\end{figure}

\begin{figure}[htb]
	\centering
		\includegraphics[width=0.5\textwidth]{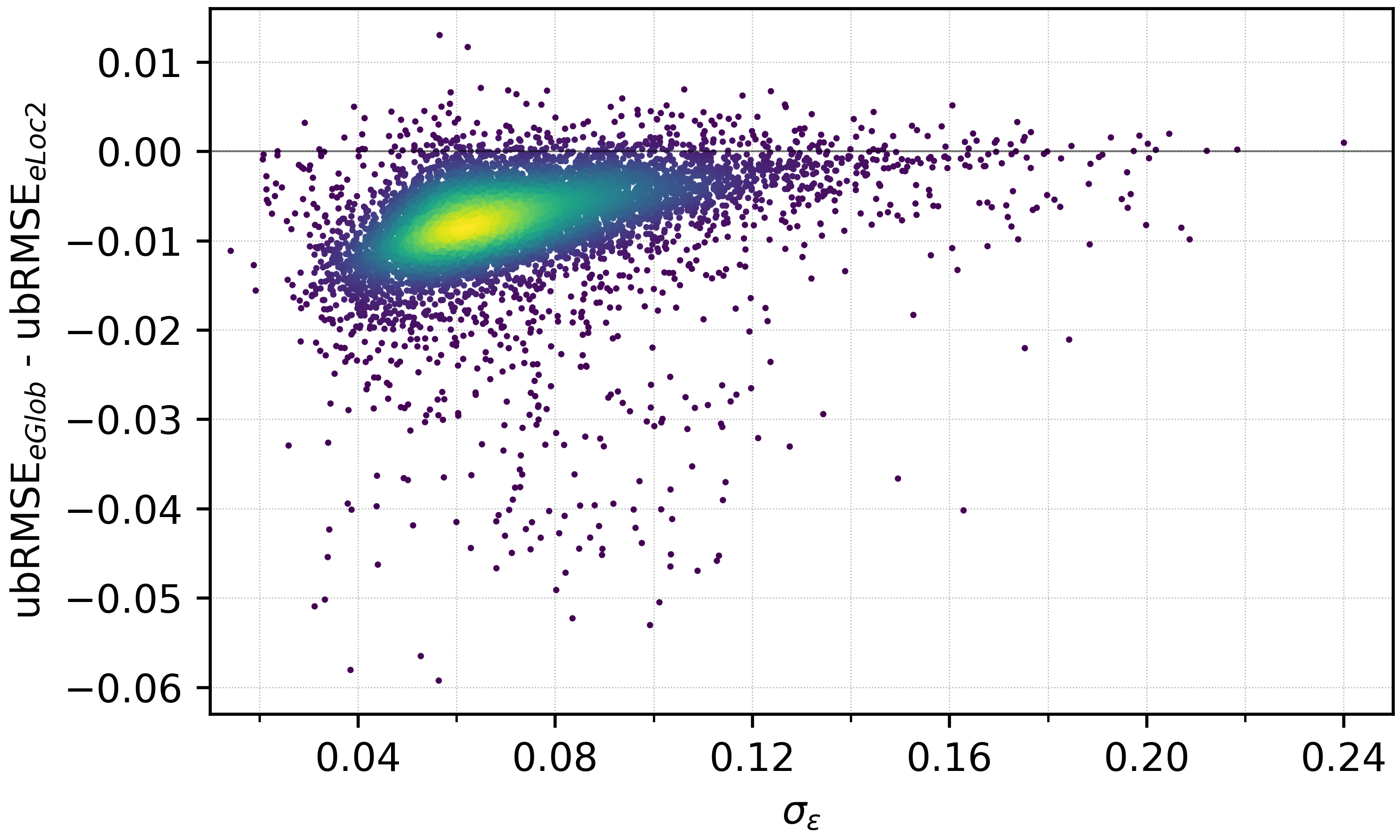}
	\caption{Difference of ubRMSE between \textit{eGlob} and \textit{eLoc2} vs. observation error obtained with TCA (average of the considered soil layers). The point colours represent the point density (yellow for high, dark purple for low). Like the ubRMSE, $\sigma_\epsilon$ is given in units of fractional SWI.
    } \label{fig:wbo}
\end{figure}

The ubRMSE adopts values that show that the water balance is a suitable measure for the intercomparison of the assimilation experiments, e.g., for \textit{eGlob}, the ubRMSE is approximately 0.08 on average. The largest ubRMSE values are found for the highest elevations where the determination of the water balance as well as of both modelled and measured soil moisture is difficult (Fig.~\ref{fig:wbmt}, left). Similarly, the correlation coefficient is significantly positive everywhere except for the highest elevations (Fig.~\ref{fig:wbmt}, right).

Averaged over the whole domain, \textit{eGlob} and \textit{eLoc1} exhibit almost identical skill metrics (Table~\ref{tab:wb}). On the other hand, small local ubRMSE differences are visible for \textit{eLoc1} (Fig.~\ref{fig:wbdm}, left). In contrast, all average skill metrics show a small degradation for \textit{eLoc2} compared to \textit{eGlob} and the spatial distribution exhibits a degradation of the ubRMSE in most of the assimilated domain except for WG2 (Fig.~\ref{fig:wbdm}, right).

The spatial pattern of the ubRMSE difference displays a small scale variability in regions without assimilation (e.g. in the Alps). Those differences arise from the spatial propagation of information through the cycled atmospheric model when using a different initialisation for the soil.

The individual monthly averages over the whole domain and all soil layers show a negligible temporal variation for the ubRMSE (Fig.~\ref{fig:wbm}). The correlation displays larger values for April and September and remains mostly constant in between.
The averaged ubRMSE difference between \textit{eGlob} and \textit{eLoc2} remains constant and slightly negative for all months. 
The averaged correlation difference is positive and exhibits a small increase towards the end of the investigated time period.
The trend here does not follow the trend of the NWP verification (Fig.~\ref{fig:vrfm}), which might be due to the different physical processes reflected by the respective reference measures. 

Similar to Sect.~\ref{sec:res}\ref{sec:avrf}, we test for land cover type dependencies of the performance metrics. Only cover types present at more than 150 model grid cells are considered for evaluation. 
In the following, we will focus on the evaluation of the ubRMSE. The correlation coefficient produces qualitatively similar results.

The ubRMSE distributions of all cover types show a general improvement of the soil moisture analysis compared to the reference run and partially a slight degradation compared to the observations (Fig.~\ref{fig:wbb}). 
In addition, the ubRMSE is different for each cover type. The largest values are found for bare land, indicating that water balance and soil moisture analysis represent very different information for this cover type. Indeed, bare land in Austria is only found on high altitudes where no SCATSAR-SWI is available and where surface models suffer in general under difficulties related to the topography \citep{2018hacdra}. The comparatively bad performance obtained for herbaceous/shrubs is unexpected though, as the evapotranspiration needed for the computation for the water balance is calibrated to grasslands \citep{1977doopru}. This could indicate that the algorithms involved in creating the input components for the data assimilation (model, observation, forcing) rely on very different assumptions for this cover type.

Nevertheless, the good agreement between water balance and soil moisture analysis (ubRMSE $<0.08$) for the remaining cover types demonstrates the suitability of the water balance as a reference measure. 

The difference between the distributions of \textit{eGlob} and \textit{eLoc1} remains in general very small for each cover type. Only the most abundant cover types crops and forests show a significant, but barely visible change for \textit{eLoc1}.
For \textit{eLoc2}, the ubRMSE degrades significantly compared to \textit{eGlob} for all cover types except for bare ground. 

To test dependencies on other surface variables, we investigate the behaviour of ubRMSE differences between the experiments compared to elevation, topographic complexity, and observation error analogous to Sect.~\ref{sec:res}\ref{sec:avrf}. No dependencies were found when comparing \textit{eGlob} and \textit{eLoc1}. When comparing \textit{eLoc2} with \textit{eGlob}, a seemingly constant degradation of around 1\% is visible for both elevation and topographic complexity (not shown).
This degradation is found to depend on the observation error, where the ubRMSE shows a degradation towards smaller errors (Fig.~\ref{fig:wbo}). That means that by putting more weight on the assimilated observations, the soil moisture analysis reproduces the water balance actually worse than with less weight (larger observation error). This behaviour is consistent for the whole investigated time period as well as for the single months.

Since the assimilated observations have on average a good agreement with the water balance (cf. Fig.~\ref{fig:wbb}), this result seems to stand in contrast to the outcome of the assimilation statistics, where the ratio of diagnosed and specified error is on average the best for \textit{eLoc2} (Sect.~\ref{sec:res}\ref{sec:astat}, Table~\ref{tab:astat}). On the one hand, a general overweighting of the observations because of neglected cross-layer error correlations is possible. By deriving the values of the profile layers from the surface soil moisture with the exponential filter, the errors cannot be assumed to be uncorrelated between the layers. That means that neglecting the off-diagonals in the error covariance matrix could have led to an unfortunate weighting of the single layers with respect to each other and, thus, to a degradation of the soil moisture analysis as described before.
On the other hand, the verification of the atmospheric forecast showed that the local observation error is able to improve the skill metrics of the forecast partially (mainly for low-elevation stations). Other studies noted already that improvements of screen-level variables are not always directly related to improvement of soil moisture fields due to the model errors \citep{2007druvit,2019faide}. We suspect that the NWP model, which is used for the atmospheric forcing of SODA, is tuned with coarser spatial scales than the one provided by the SCATSAR-SWI and, therefore, prevents the soil moisture analysis from profiting from the supposedly better local error specification.

\section{Conclusion} \label{sec:sum}

We assimilated the multi-layer soil moisture product SCATSAR-SWI into the AROME/SURFEX NWP system using an sEKF in the Austrian domain. 
With the goal to improve the specification of the Kalman gain, observation error variances were determined locally with TCA, using SMAP and model-based SURFEX soil moisture as additional data sets.

The assimilation system was run in three configurations, one using a global observation error (\textit{eGlob}) and two using the local observation errors obtained with TCA (\textit{eLoc1}, \textit{eLoc2}). The \textit{eLoc1} specification utilised a scaled version of the TCA error distribution to adopt the same average as the \textit{eGlob} errors, while \textit{eLoc2} employed the TCA values without modification. 
For \textit{eLoc1}, the DA system stayed close to \textit{eGlob} for all tested measures. The results for \textit{eLoc2} were more mixed as described below.

When evaluating the performance of the data assimilation system, the analysis of the Jacobians of the observation operator showed that a specific soil layer is generally more sensitive to the upper than to the lower layer control variables. Sensitivities can show larger differences between regions with different topographies.

Innovation and residual statistics as well as the resulting increments revealed that \textit{eLoc2} might have wettened the soil too much on average. Nevertheless, the ratio of specified to diagnosed errors improved compared to the other assimilation experiments.

The verification of the NWP forecast against Austrian weather stations showed that the assimilation runs partially improved the skill metrics for both 2~m temperature and 2~m relative humidity when comparing against the reference run. A large improvement was found for the night time performance, where the more humid soil moisture analysis proved beneficial for initialising the NWP model. Large average differences were also found for the RMSE in July and August, presumably because of the potential annual maximum of soil-atmosphere interactions at that time, which allows for a strong impact of the changes in soil moisture onto the NWP. 

Dependencies on different parameters, such as station altitude, land cover type, or magnitude of observation error were investigated. Forecasts for stations at lower elevations were found to benefit the most from the local error specification (\textit{eLoc2}).

Using precipitation and evapotranspiration fields, we created a water balance approximation to obtain a direct reference measure for the soil moisture analysis. The absolute comparability of soil moisture with the water balance is non-trivial due to the simplifications needed for computing the water balance and for the ingested computation of the evapotranspiration. Nevertheless, the water balance proved useful for the intercomparison of different assimilation experiments.

We found that \textit{eLoc2} degrades the ubRMSE on large scales compared to \textit{eGlob}. To probe the small-scale variability, we tested for dependencies on different surface parameters. The performance metrics revealed a general dependency on the land cover type. Most strikingly, \textit{eLoc2} exhibited a degradation of the ubRMSE with decreasing specified observation errors. This effect was tentatively attributed to the tuning of the NWP model and a potential overweighting of single soil layers.

To summarise, the approach of using TCA to compute local error variances seems promising to assess the general quality of the data. The implementation of the TCA errors in the soil moisture assimilation yielded partially positive results, which can help to better understand the assimilation system and its components in general. Nevertheless, the evaluation of the observation errors based on the investigated surface variables did not show a benefit for the soil moisture analysis. As a consequence, we tend to the conclusion that the effort involved in applying TCA is currently not worthwhile for operational applications, especially regarding the necessity of having three independent data sets. 
More effort, especially with regard to error cross correlations and the configuration of the NWP model, might be necessary to fully exploit the TCA approach.

\acknowledgments
This research was performed in the framework of a EUMETSAT fellowship for the project EHRSOMA (European High-Resolution Soil Moisture Analysis).
We thank Alexander Gruber for his valuable comments, which helped to improve this paper.
In addition, we would like to thank the anonymous referees for their important input.

\datastatement
The SCATSAR-SWI is openly available on the Copernicus Global Land Service \url{https://land.copernicus.vgt.vito.be/PDF/portal/Application.html}. The SMAP data is available on \url{https://nsidc.org/data/smap/smap-data.html}. Due to its proprietary nature, the measurements of the weather stations and the data needed for creating the water balance product cannot be made openly available. Further information about the data and conditions for access are available at \url{https://www.zamg.ac.at/cms/en/climate/
meteorological-network} and \url{https://www.zamg.ac.at/cms/de/forschung/klima/klimatografien/spartacus}.

\bibliographystyle{ametsoc2014}
\bibliography{BIBB}

\end{document}